\documentstyle{mn}

\newif\ifAMStwofonts \AMStwofontstrue

\def\h2o{{$\mbox{H}_{2}\mbox{O}$}} \def\kms{{$\mbox{km~s}^{-1}$}}
\def\nat{Nature} \def\apj{ApJ} \def\apjl{ApJ} \def\apjs{ApJS}
\def\aj{AJ} \def\mnras{MNRAS} \def\aap{A\&A} \def\pasj{PASJ}
\def\pasa{PASA} \def\iaucirc{IAU~Circ.}

\newcommand{\wsf}{1}    
\newcommand{\wsv}{2}    
\newcommand{\lss}{3}    
\newcommand{\dsn}{4}    
\newcommand{\bet}{5}    
\newcommand{\get}{6}    
\newcommand{\bef}{7}    
\newcommand{\nef}{8}    
\newcommand{\mef}{9}    
\newcommand{\cef}{10}   
\newcommand{\nev}{11}   
\newcommand{\neva}{12}  
\newcommand{\tev}{13}   
\newcommand{\bgev}{14}  
\newcommand{\hex}{15}   
\newcommand{\uex}{16}   
\newcommand{\nex}{17}   
\newcommand{\cex}{18}   
\newcommand{\wex}{19}   
\newcommand{\ccex}{20}  
\newcommand{\bex}{21}   
\newcommand{\ssex}{22}  
\newcommand{\bgses}{23} 
\newcommand{\sses}{24}  
\newcommand{\bhes}{25}  
\newcommand{\bhhes}{26} 
\newcommand{\hmes}{27}  
\newcommand{\hes}{28}   
\newcommand{\nes}{29}   
\newcommand{\mes}{30}   
\newcommand{\bges}{31}  
\newcommand{\mee}{32}   
\newcommand{\nee}{33}   
\newcommand{\kee}{34}   
\newcommand{\meea}{35}  
\newcommand{\meeb}{36}  
\newcommand{\ssen}{37}  
\newcommand{\ben}{38}   
\newcommand{\den}{39}   
\newcommand{\nen}{40}   
\newcommand{\men}{41}   
\newcommand{\bhen}{42}  
\newcommand{\bgen}{43}  
\newcommand{\ken}{44}   
\newcommand{\cno}{45}   
\newcommand{\hno}{46}   
\newcommand{\kno}{47}   
\newcommand{\dno}{48}   
\newcommand{\knl}{49}   
\newcommand{\mnt}{50}   
\newcommand{\bnt}{51}   
\newcommand{\ssnt}{52}  
\newcommand{\wnt}{53}   
\newcommand{\bnf}{54}   
\newcommand{\lnf}{55}   
\newcommand{\enf}{56}   
\newcommand{\vlnv}{57}  
\newcommand{\rnv}{58}   
\newcommand{\nnv}{59}   
\newcommand{\gnx}{60}   
\newcommand{\bnx}{61}   
\newcommand{\ggnx}{62}  
\newcommand{\knx}{63}   
\newcommand{\gns}{64}   

\ifoldfss
\ifCUPmtlplainloaded \else \NewTextAlphabet{textbfit} {cmbxti10} {}
\NewTextAlphabet{textbfss} {cmssbx10} {}
    \NewMathAlphabet{mathbfit} {cmbxti10} {} 
    \NewMathAlphabet{mathbfss} {cmssbx10} {} 
    \fi
    \ifAMStwofonts
    \ifCUPmtlplainloaded \else \NewSymbolFont{upmath} {eurm10}
    \NewSymbolFont{AMSa} {msam10} \NewMathSymbol{\upi} {0}{upmath}{19}
    \NewMathSymbol{\umu} {0}{upmath}{16}
    \NewMathSymbol{\upartial}{0}{upmath}{40}
    \NewMathSymbol{\leqslant}{3}{AMSa}{36}
    \NewMathSymbol{\geqslant}{3}{AMSa}{3E}

     \fi
    \fi
\fi 

\ifnfssone
\newmathalphabet{\mathit} \addtoversion{normal}{\mathit}{cmr}{m}{it}
\addtoversion{bold}{\mathit}{cmr}{bx}{it}
  \newmathalphabet{\mathbfit} 
  \addtoversion{normal}{\mathbfit}{cmr}{bx}{it}
  \addtoversion{bold}{\mathbfit}{cmr}{bx}{it}
  \newmathalphabet{\mathbfss} 
  \addtoversion{normal}{\mathbfss}{cmss}{bx}{n}
  \addtoversion{bold}{\mathbfss}{cmss}{bx}{n}
  \ifAMStwofonts
  \ifCUPmtlplainloaded \else
  \UseAMStwoboldmath
  \makeatletter \new@mathgroup\upmath@group
  \define@mathgroup\mv@normal\upmath@group{eur}{m}{n}
  \define@mathgroup\mv@bold\upmath@group{eur}{b}{n}
  \edef\UPM{\hexnumber\upmath@group}
  \new@mathgroup\amsa@group
  \define@mathgroup\mv@normal\amsa@group{msa}{m}{n}
  \define@mathgroup\mv@bold\amsa@group{msa}{m}{n}
  \edef\AMSa{\hexnumber\amsa@group} \makeatother
  \mathchardef\upi="0\UPM19 \mathchardef\umu="0\UPM16
  \mathchardef\upartial="0\UPM40 \mathchardef\leqslant="3\AMSa36
  \mathchardef\geqslant="3\AMSa3E
  \fi \fi
\fi 

\ifnfsstwo
  \DeclareMathAlphabet{\mathbfit}{OT1}{cmr}{bx}{it}
  \SetMathAlphabet\mathbfit{bold}{OT1}{cmr}{bx}{it}
  \DeclareMathAlphabet{\mathbfss}{OT1}{cmss}{bx}{n}
  \SetMathAlphabet\mathbfss{bold}{OT1}{cmss}{bx}{n}
  \ifAMStwofonts
  \ifCUPmtlplainloaded \else \DeclareSymbolFont{UPM}{U}{eur}{m}{n}
  \SetSymbolFont{UPM}{bold}{U}{eur}{b}{n}
  \DeclareSymbolFont{AMSa}{U}{msa}{m}{n}
  \DeclareMathSymbol{\upi}{0}{UPM}{"19}
  \DeclareMathSymbol{\umu}{0}{UPM}{"16}
  \DeclareMathSymbol{\upartial}{0}{UPM}{"40}
  \DeclareMathSymbol{\leqslant}{3}{AMSa}{"36}
  \DeclareMathSymbol{\geqslant}{3}{AMSa}{"3E}
  \fi \fi
\fi 

\ifCUPmtlplainloaded \else
  \ifAMStwofonts \else 
  \def\upi{\pi} \def\umu{\mu} \def\upartial{\partial} \fi \fi

  \title{A comprehensive search for extragalactic 6.7-GHz methanol
    masers} \author[C.J.~Phillips et al.]  {C.J.~Phillips,$^1$
    R.P.~Norris,$^2$ S.P.~Ellingsen$^1$ and
    D.P.~Rayner$^1$\\
    $^1$Department of Physics, University of Tasmania, GPO Box 252-21,
    Hobart, TAS 7001, Australia.\\
    $^2$CSIRO Australia Telescope National Facility, PO Box 76,
    Epping, NSW 2121, Australia.}
  
  \date{Accepted 19xx.  Received 19xx}
  
  \pagerange{\pageref{firstpage}--\pageref{lastpage}} \pubyear{19xx}

\begin{document}

\maketitle

\label{firstpage}

\begin{abstract}
  We have used the Australia Telescope Compact Array (ATCA) to search
  for 6.7-GHz methanol maser emission towards 87 galaxies. We chose
  the target sources using several criteria, including far-IR
  luminosities and the presence of known OH megamasers. In addition,
  we searched for methanol masers in the nearby starburst galaxy NGC
  253, making a full spectral-line synthesis image. No emission was
  detected in any galaxies, with detection limits ranging from 25~mJy
  to 75~mJy. This is surprising, given the close association of OH and
  methanol masers in Galactic star-formation regions, and
  significantly constrains models of OH megamaser emission. This
  absence of maser emission may be a result of low methanol abundances
  in molecular clouds in starburst galaxies.

\end{abstract}


\section{Introduction}

Extragalactic maser emission was first discovered in NGC~253 in the
1665- and 1667-MHz transitions of OH \cite{wg74} with a luminosity
about 100 times greater than typical Galactic OH masers. In 1982 an OH
``megamaser'', a maser a million times more luminous than the most
luminous Galactic OH masers, was discovered in Arp220 (IC 4553) (Baan,
Wood \& Haschick 1982). Since then more than 50 extragalactic OH
sources (often collectively called megamasers, although some of them
extend to lower luminosities) have been discovered \cite{b93}. These
OH sources are characterised by broad emission lines (up to several
100 \kms) and stronger emission at 1667~MHz than at 1665~MHz. This is
in contrast to Galactic OH masers in star-forming regions which have a
typical velocity range of 10 \kms\ and emission stronger at 1665~MHz
than at 1667~MHz.

The standard model of OH megamasers (typified by Arp220) is that
low-gain amplification occurs in a molecular disc around the nucleus
of the galaxy \cite{n84}. The gas is pumped by far-infrared radiation
\cite{b89} and amplifies the nuclear continuum emission.

The number of known extragalactic \h2o\ masers is smaller than the
number of OH masers, and can be divided into two types. The first
extragalactic \h2o\ maser was discovered in M33 \cite{cw77}. This has
a luminosity similar to the brightest Galactic maser, W49.  Over a
dozen \h2o\ masers of a similar luminosity have been discovered in
other galaxies, including NGC~253. In 1979 a new class of \h2o\ masers
was discovered in NGC~4945 \cite{dl79}.  The intrinsic brightness of
NGC~4945 is about 100 times greater than W49. 15 similarly bright
masers have since been discovered, and the strongest, TXFS2226-184, is
5000 times more luminous than W49, and seven orders of magnitude
stronger than typical Galactic \h2o\ masers \cite{kh95}. These \h2o
megamasers occur within a few pc of the galactic nucleus
\cite{cl86,gm90}. In the Claussen \& Lo model the maser emission occurs
in a dense disc of gas and dust close to the nucleus of the galaxy.
The masers are believed to be collisionally pumped as a result of the
gas being heated by X-rays \cite{nm94,nm95}.

OH and \h2o\/ megamasers are extremely powerful astrophysical tools.
For example, VLBI observations of the \h2o\ megamasers in NGC~4258 have
given the strongest evidence yet for the existence of massive black
holes in active galaxies.  The masers in this galaxy clearly show a
Keplerian rotation, indicating a central mass of 3.5$\times10^7$
M$_{\sun}$ in a region less than 0.13 pc \cite{mm95,gj95}. Similar
results are being found in NGC~1068 and the Circinus galaxy
\cite{gg96,ge97}.

Strong methanol emission in the 12.2-GHz ($2_{0}-3_{-1}$~E) and
6.7-GHz ($5_1-6_0 \mbox{A}^+$) transitions is often found in regions
of star-formation \cite{bm87,m91}. VLBI observations show that the
12.2- and 6.7-GHz masers are often spatially coincident
\cite{mr92,e95}. These methanol transitions are also closely
associated with OH, often coincident to within less than 1 arcsec
(Caswell, Vaile \& Forster 1995).

Interferometric imaging of these sources has shown that they are often
located in lines or arcs, with a simple velocity gradient along the
line (Norris et al. 1993, 1997; Phillips et al. 1996). Norris et al.
\shortcite{nb97} interpret these sources as edge-on circumstellar
discs around young massive stars.

Although $\sim$400 Galactic 6.7-GHz methanol masers have been found
(Ellingsen et al. 1996 and references therein), only three
extragalactic methanol masers are known, all in the Large Magellanic
Clouds \cite{sc92,ew94,be96}. These masers are unremarkable and their
intrinsic brightness is similar to Galactic methanol masers. Only one
published search for 6.7-GHz methanol megamasers has been made, in
which ten known OH and \h2o\ megamaser galaxies were searched for
6.7-GHz emission, but none was found \cite{en94}. Given the close
association of methanol and OH in Galactic masers, this is surprising.

We present here a survey of 87 galaxies for 6.7-GHz megamaser
emission, covering a variety of different types of galaxies. We have also
done full spectral-line imaging on the edge-on starburst galaxy,
NGC~253, to detect either megamaser emission or else strong maser
emission associated with star-formation regions.

\section{Observations}
\subsection{NGC~253}
The observations were made with the Australia Telescope Compact Array
during an 11-h period on 1994 July 28  with the array in the 6A
configuration, which gives 15 baselines ranging in length from 337 to
5939 m. A bandwidth of 8~MHz was used with the correlator configured
to give 1024 spectral channels in each of two orthogonal linear
polarizations.  The typical system temperature of the antennae at
6.7-GHz is 120 K.

Two pointing centres were required to accommodate the large velocity
spread across the galaxy ($\sim$450 \kms) and its large angular size
(20 arcmin compared with a primary beam FWHM of 8.4 arcmin). The
positions used were (00:47:38.98,~-25:16:13.2) and
(00:47:27.39,~-25:18:19.1) J2000, with central frequencies 6664 and
6660~MHz, respectively. The two pointing centres were each observed 11
times for 25 min over 11 h, interspersed with observations of
the secondary calibrator 0023-263.

The data were calibrated and imaged using the Astronomical Image
Processing System (AIPS). 1934-638 was used as the primary flux
calibrator, with an assumed flux density of 3.93~Jy at 6.6 GHz.  After
the initial calibration, the spectral channels were averaged and a
continuum image made. The centre of the galaxy was clearly detected in
both images with a flux density of 220~mJy. These continuum images
were then phase self-calibrated (improving the signal-to-noise ratio
from 130 to 300) and the improved phase calibration applied to the
un-averaged spectral-line data. Because the spectral resolution of
0.35 \kms\ is comparable to a typical FWHM of 0.5 \kms\ for Galactic
masers \cite{ev96}, the data were not Hanning smoothed.  Two
825-channel spectral cubes (with a velocity span of 290 \kms) were
made for each of the pointing centres.  Each cube covered
256$\times$256 arcsec with an angular resolution of 2.9$\times$1.3
arcsec.  Together the two cubes covered approximately the inner third
of the galaxy.

\subsection{Megamaser survey}

Since methanol is likely to be abundant only in molecular-rich
galaxies, which are typically {\em IRAS} sources, the source selection
for this survey included only southern {\em IRAS} galaxies, with known
redshifts, known 60~$\mu$m flux density, and declination $<$~+20
(except Arp 220). No sources with z~$>$~0.15 were considered.
Preference was given to sources that satisfied one or more of the
following criteria:
\begin{enumerate}
\item known OH megamaser galaxy
\item known \h2o\ megamaser galaxy
\item 60~$\mu$m flux (S60) $>$ 10 Jy
\item S60 (Jy) $\times$ z $>$ 0.3
\item log($\mbox{L}_{\sun}$) $>$ 12.
\end{enumerate}
The observations were made between 1996 May 28 and 1996 June 3 using
the ATCA, with the array in the 750D configuration, with 10 baselines
ranging in length from 30.6 to 719.4 m. A bandwidth of 16~MHz was used
with the correlator configured to give 512 channels for two orthogonal
linear polarizations.

87 sources were observed from the sample discussed above.
Table~\ref{mmmtab} lists the observed sources and their selection
criteria. Each source was observed for 1 h, with an observation
of a secondary calibrator made before and after each source. Sources
were observed as close to transit as possible to simplify data
processing (as discussed below), although this meant some of the
observed sources did not satisfy any of the selection criteria
specified above (because of the uneven distribution of the sources on
the sky).

There are several advantages in using an interferometer in stead of a
single dish for spectral-line detection experiments.

$\bullet$ Single-dish observations are often made in
``position-switch'' mode, spending half the observing time at a
reference position and creating a ``quotient spectrum''. This is not
required for an interferometer, enabling the same sensitivity to be
achieved in one quarter of the time.

$\bullet$ Interferometers are also less susceptible to variations in
the bandpass, enabling longer integrations.

The disadvantages are that neither a scalar nor a vector average of
interferometer data produces the same response as a single dish.

$\bullet$ A scalar average spectrum of the data (averaging the
individual power spectra from each baseline and each integration
period) produces a lower sensitivity to weak spectral features because
of noise bias \cite{tm86}, which appears as a positive baseline offset
in the spectrum, but under which weak spectral features are submerged.

$\bullet$ A vector average of the data (adding the real and imaginary
parts separately, after phase rotation to a suitable phase centre
position) overcomes this problem, but the spectrum is restricted to an
area the size of the synthesised beam of the interferometer, centred
at the averaging phase-centre.

For many of the sources in the sample, the position of the centre of
the Galaxy was not sufficiently well known to guarantee that the maser
would lie at the phase centre of the synthesised beam. Making spectral
cubes of all sources observed was not practical and so, instead, we
overcame this problem by making all observations close to transit, so
that the synthesised beam was approximately
15 arcsec $\times$ 2 arcmin, elongated in a north-south direction. We
then plotted a series of spectra each with a different phase centre.
By making a series of seven plots with the phase centre shifted by
10~arcsec from east to west across the observed position, we are
sensitive to emission from sources within $\pm$30~arcsec of the
nominal galaxy position.

For normal interferometric observations the antenna-based gains and
delays need to be determined before each observing session, by
observing both a primary and secondary flux calibrator at each
observing frequency. To maximize the efficiency of the search, we
chose not to do a full amplitude calibration, but instead estimated
gains based on the assumed secondary calibrator flux density, with
additional calibration observations planned in the event of a
detection. Since only rough delay calibration was used, we expect a
10-20 percent error caused by decorrelation during the observation. The
amplitude calibration for most sources should be better than about
30 percent for most sources, though some may be only within a factor of two.

Basic calibration of the data was done using the AIPS software.  A
series of seven spectra, shifted in position as described above, was
plotted for each source. To facilitate the detection of both narrow
and broad emission, plots were produced with no smoothing, Hanning
smoothing, and boxcar smoothing over 5 and 10 channels.  The spectral
resolution of the four smoothing schemes was 1.5, 3.0, 7.4 and 14.9
\kms\ respectively.

\section{Results}
\subsection{NGC~253}

The rms noise was typically 23~mJy in each 0.35~\kms channel of the
two image cubes. The peaks in each cube were found (the largest $\sim$
130~mJy) and carefully checked for real emission in the image cube and
u-v data. We did not detect any methanol emission. Plots of the
spatial distribution of the peaks in the cube were consistent with
noise, and there was no correlation with the position of the galaxy.
The noise profile was nearly Gaussian but with slightly broader wings.
Assuming a 5-$\sigma$ detection threshold implies no 6.7-GHz methanol
stronger than 107~mJy.  Assuming a distance of 3.4 Mpc to NGC~253
\cite{h72}, the most luminous Galactic 6.7-GHz methanol maser,
G340.78-0.10 \cite{nw93}, would have a flux density of 2.6~mJy in
NGC~253. Thus our observations were not sensitive enough to detect
Galactic-strength methanol masers in NGC 253, but would have detected
masers 35 times stronger. The OH masers in NGC~253 are about 100 times
stronger than those in our Galaxy, and so a corresponding experiment
at OH wavelengths would have detected the masers at the 14-$\sigma$
level.

\begin{table*}
 \centering
 \begin{minipage}{180mm}
 \caption{The selected sample of galaxies}
 \label{mmmtab}
\begin{tabular}{ll r@{:}c@{:}l r@{:}c@{:}l rrlr@{.}llrl}

\multicolumn{1}{c}{\em IRAS} & & \multicolumn{6}{c}{\bf Position (J2000)} & 
 \multicolumn{1}{c}{\bf OH} & \multicolumn{1}{c}{\bf \h2o} & 
 \multicolumn{1}{c}{} & \multicolumn{2}{c}{\em IRAS} & 
 \multicolumn{1}{c}{\bf Sel.$^a$} & \multicolumn{1}{c}{\bf RMS$^b$} & \\

\multicolumn{1}{c}{\bf Name}  & \multicolumn{1}{l}{\bf Alias} & 
 \multicolumn{3}{c}{\bf RA} & \multicolumn{3}{c}{\bf Dec} & 
 \multicolumn{2}{c}{\bf Peak Flux} & \multicolumn{1}{c}{\bf z} & 
 \multicolumn{2}{c}{\bf 60$\mu$m} & \multicolumn{1}{c}{\bf Crit.} & 
 \multicolumn{1}{c}{\bf Noise} & {\bf References$^{c}$} \\

& & \multicolumn{3}{c}{} & \multicolumn{3}{c}{} & 
 \multicolumn{2}{c}{\bf (mJy)} & \multicolumn{1}{c}{} & 
 \multicolumn{2}{c}{\bf Jy} & & \multicolumn{1}{c}{\bf (mJy)} & \\
\\
00198$-$7926 &             & $00$&$21$&$54$ & $-79$&$10$&$08$ &       &         & 0.0724 &   3&2  & o &   14.3 & \\
00335$-$2732 &             & $00$&$36$&$01$ & $-27$&$15$&$34$ & $<$20 &         & 0.0691 &   4&4  & z &    9.1 & \tiny{\ssnt,\wnt} \\
00450$-$2533 & NGC~253     & $00$&$47$&$33$ & $-25$&$17$&$18$ &   120 &    5000 & 0.001  & 759&0  &hszw&  16.6 & \tiny{\wsf,\lss,\tev,\hex,\uex,\hmes,\nes} \\
\multicolumn{15}{c}{}   &                                                                                      \tiny{\nee,\ben,\hno,\ssnt,\enf,\nnv} \\
01418$+$1651 & III Zw 35   & $01$&$44$&$31$ & $+17$&$06$&$09$ &   240 &         & 0.0273 &  11&9  &hsz&   19.0 & \tiny{\ccex,\sses,\nes,\mes,\ben,\bhen,\cno} \\
\multicolumn{15}{c}{}   &                                                                                      \tiny{\hno,\mnt,\wnt,\rnv,\knx} \\
01458$-$2828 &             & $01$&$48$&$09$ & $-28$&$14$&$02$ &       &         & 0.1352 &   0&8  & o &   19.7 & \\
02401$-$0013$^{d}$&NGC~1068& $02$&$42$&$41$ & $-00$&$00$&$48$ &     5 &     670 & 0.0038 & 186&0  & o &   30.0 & \tiny{\cef,\hex,\cex,\enf,\nnv,\gnx,\bnx,\ggnx} \\
02512$+$1446 & Z 440-30    & $02$&$54$&$02$ & $+14$&$58$&$15$ &       &         & 0.031  &   7&7  & o &   21.0 & \\
03317$-$3618 & NGC~1365    & $03$&$33$&$36$ & $-36$&$08$&$23$ &    50 &  $<$210 & 0.0055 &  78&2  &hsz&   22.8 & \tiny{\nen,\nnv,\bnx} \\
03359$+$1523 &             & $03$&$38$&$47$ & $+15$&$32$&$54$ &       &         & 0.035  &   5&8  & o &   18.5 & \\
03540$-$4230$^{d}$&NGC~1487& $03$&$55$&$45$ & $-42$&$22$&$07$ &       &         & 0.0026 &   3&3  & o &    8.0 & \tiny{\enf} \\
04189$-$5503 & NGC~1566    & $04$&$20$&$00$ & $-54$&$56$&$18$ & $<$32 &  $<$340 & 0.005  &  12&7  & s &   25.2 & \tiny{\nen,\enf,\bnx} \\
04191$-$1855 & ESO 550-25  & $04$&$21$&$20$ & $-18$&$48$&$39$ &       &         & 0.0308 &   5&8  & o &   25.3 & \\
05059$-$3734 & NGC~1808    & $05$&$07$&$42$ & $-37$&$30$&$46$ & $<$32 &  $<$300 & 0.0033 &  97&1  & sz&   34.2 & \tiny{\nen,\ssnt,\nnv,\bnx} \\
05100$-$2425 &             & $05$&$12$&$09$ & $-24$&$21$&$54$ &    18 &         & 0.0338 &   4&1  & h &   13.6 & \tiny{\hno,\ssnt} \\
05189$-$2524 &             & $05$&$21$&$01$ & $-25$&$21$&$45$ &    30 &         & 0.0415 &  13&9  &hsz&   14.6 & \tiny{\sses,\nen,\ssnt} \\
05238$-$4602 & ESO 253-3   & $05$&$25$&$18$ & $-46$&$00$&$18$ &       &         & 0.0407 &   2&8  & o &    1.9 & \\
06035$-$7102 &             & $06$&$02$&$54$ & $-71$&$03$&$12$ & $<$30 &         & 0.0796 &   5&0  & lz&   11.4 & \tiny{\ssnt} \\
06076$-$2139 &             & $06$&$09$&$45$ & $-21$&$40$&$22$ & $<$15 &         & 0.0374 &   6&3  & o &   10.8 & \tiny{\ssnt} \\
06102$-$2949 & ESO 425-13  & $06$&$12$&$12$ & $-29$&$50$&$31$ & $<$40 &         & 0.0611 &   4&4  & o &   14.5 & \tiny{\ssnt} \\
06206$-$6315 &             & $06$&$21$&$01$ & $-63$&$17$&$23$ & $<$20 &         & 0.0917 &   4&0  & lz&    9.3 & \tiny{\ssnt} \\
06219$-$4330 &             & $06$&$23$&$25$ & $-43$&$31$&$45$ & $<$25 &         & 0.063  &   4&7  & o &    8.1 & \tiny{\ssnt} \\
06259$-$4708 & ESO 255-7   & $06$&$27$&$22$ & $-47$&$10$&$37$ & $<$25 &         & 0.038  &   9&4  & z &    8.7 & \tiny{\ssnt} \\
08014$+$0515 & Mrk 1210    & $08$&$04$&$06$ & $+05$&$06$&$50$ &       &     160 & 0.013  &   1&8  & w &   22.7 & \tiny{\bnf,\bnx} \\
08071$+$0509 &             & $08$&$09$&$48$ & $+05$&$01$&$03$ & y$^e$ &         & 0.0543 &   4&8  & h &    9.3 & \tiny{\bgen} \\
08520$-$6850 &             & $08$&$52$&$32$ & $-69$&$01$&$54$ & $<$20 &         & 0.0463 &   5&8  & o &   23.8 & \tiny{\ssnt} \\
09004$-$2031 & ESO 64-11   & $09$&$02$&$46$ & $-20$&$43$&$30$ & $<$15 &         & 0.0085 &   8&7  & o &   96.2 & \tiny{\sses,\ssnt} \\
09149$-$6206 & QSO0914-621 & $09$&$16$&$09$ & $-62$&$19$&$29$ &       &         & 0.0573 &   2&5  & o &   15.6 & \\
10039$-$3338 & ESO 374-32  & $10$&$06$&$07$ & $-33$&$53$&$22$ &   240 &         & 0.0337 &   9&1  &hz &   18.0 & \tiny{\kee,\hno,\kno,\ssnt,\enf,\rnv,\knx} \\
10173$+$0828 &             & $10$&$20$&$00$ & $+08$&$13$&$34$ &   100 &         & 0.0485 &   6&1  & h &   21.9 & \tiny{\mes,\ben,\hno} \\
10221$-$2317 & ESO 500-34  & $10$&$24$&$31$ & $-23$&$33$&$12$ & $<$70 &         & 0.013  &  11&3  & s &   27.0 & \tiny{\nen} \\
10257$-$4338 & NGC~3256    & $10$&$27$&$52$ & $-43$&$54$&$09$ & $<$30 &         & 0.0096 &  94&6  & sz&   78.8 & \tiny{\nen,\ssnt} \\
11095$-$0238 &             & $11$&$12$&$03$ & $-02$&$54$&$18$ &       &         & 0.106  &   3&2  & lz&   12.7 & \\
11143$-$7556 &             & $11$&$16$&$04$ & $-76$&$12$&$53$ & $<$20 &         & 0.0054 &  47&7  & s &   48.4 & \tiny{\nen,\ssnt} \\
11365$-$3727 & NGC~3783    & $11$&$36$&$33$ & $-37$&$27$&$42$ &       &  $<$250 & 0.0107 &   3&4  & o &   23.6 & \tiny{\bnx} \\
11506$-$3851 & ESO 320-30  & $11$&$53$&$12$ & $-39$&$07$&$49$ &    90 &  $<$850 & 0.010  &  34&2  &hsz&   22.2 & \tiny{\nex,\nes,\ben,\nen,\hno,\ssnt,\wnt,\enf,\bnx} \\
11581$-$2033 & ISZ 096     & $12$&$00$&$43$ & $-20$&$50$&$07$ &       &         & 0.0621 &   1&7  & o &   11.9 & \\
12112$+$0305 &             & $12$&$13$&$46$ & $+02$&$48$&$41$ &    45 &         & 0.073  &   8&4  &hlz&   12.9 & \tiny{\mee,\ben,\bhen,\hno,\knl,\bnt,\wnt} \\
12232$+$1256 & NGC~4388    & $12$&$25$&$47$ & $+12$&$39$&$41$ & $<$20 &   $<$60 & 0.0085 &  10&9  & s &   25.7 & \tiny{\uex,\bnt,\bnx} \\
12243$-$0036 & NGC~4418    & $12$&$26$&$57$ & $-00$&$52$&$50$ &     7 &  $<$125 & 0.007  &  43&9  &hsz&   29.3 & \tiny{\sses,\nes,\bges,\meeb,\ben,\nen,\hno} \\
\multicolumn{15}{c}{}   &                                                                                        \tiny{\wnt,\enf,\rnv,\bnx} \\
12294$+$1441 & NGC~4501    & $12$&$31$&$59$ & $+14$&$25$&$10$ &       &  $<$160 & 0.0077 &  14&2  & s &   27.6 & \tiny{\hex,\bnx} \\
13001$-$2339 & ESO507-70   & $13$&$02$&$52$ & $-23$&$55$&$17$ & $<$18 &         & 0.0209 &  14&1  & s &   26.3 & \tiny{\nen,\ssnt} \\
13025$-$4911 & NGC~4945    & $13$&$05$&$26$ & $-49$&$28$&$16$ &       &    6200 & 0.0020 & 388&1  &wsz&  160.7 & \tiny{\wsf,\wsv,\dsn,\mef,\nes,\hno,\ssnt,\nnv,\bnx} \\
13225$-$4245 & Centaurus A & $13$&$25$&$28$ & $-43$&$01$&$09$ &   120 &  $<$580 & 0.0018 & 171&0  &hsz&  645.5 & \tiny{\nes,\vlnv,\bnx} \\
13335$-$2612 &             & $13$&$36$&$22$ & $-26$&$27$&$30$ &       &         & 0.1248 &   1&4  & o &   15.5 & \\
13451$+$1232 &             & $13$&$47$&$33$ & $+12$&$17$&$24$ &   1.7 &         & 0.122  &   2&0  & h &   45.1 & \tiny{\dno} \\
14092$-$6506 & Circinus    & $14$&$13$&$10$ & $-65$&$20$&$22$ &       &   16000 & 0.0013 & 248&7  &wsz&   56.3 & \tiny{\get,\mef,\wex,\nes,\enf,\nnv,\bnx,\gns} \\ 
14147$-$2248 &             & $14$&$17$&$36$ & $-23$&$02$&$02$ & $<$25 &         & 0.0794 &   2&4  & o &   11.2 & \tiny{\ssnt} \\
14348$-$1447 & GNH 35      & $14$&$37$&$38$ & $-15$&$00$&$24$ &       &         & 0.0824 &   6&8  & lz&   16.6 & \\
14376$-$0004 & NGC~5713    & $14$&$40$&$11$ & $-00$&$17$&$20$ & $<$12 &         & 0.007  &  19&8  & s &   41.1 & \tiny{\uex,\bnt} \\
14378$-$3651 &             & $14$&$40$&$58$ & $-37$&$04$&$25$ & $<$20 &         & 0.0682 &   6&5  & lz&   11.7 & \tiny{\ssnt} \\
15107$+$0724 & Zw 049.057  & $15$&$13$&$13$ & $+07$&$13$&$35$ &    13 &         & 0.012  &  21&6  &hs &   19.3 & \tiny{\bex,\bhhes,\meea,\ben,\hno,\wnt} \\
15247$-$0945 &             & $15$&$27$&$26$ & $-09$&$55$&$57$ &  y$^e$&         & 0.0400 &   4&7  & h &   13.2 & \tiny{\ken,\hno} \\
15268$-$7757 & ESO 22-10   & $15$&$33$&$36$ & $-78$&$07$&$28$ & $<$25 &         & 0.0087 &   4&2  & o &   16.5 & \tiny{\ssnt} \\
15322$+$1521 & NGC~5953    & $15$&$34$&$32$ & $+15$&$11$&$42$ &       &   $<$90 & 0.0066 &  10&4  & s &   20.0 & \tiny{\bnx} \\
15327$+$2340 & Arp220      & $15$&$34$&$57$ & $+23$&$30$&$12$ &   300 &  $<$200 & 0.0182 & 104&0  &hsz&   28.0 & \tiny{\bet,\bef,\nef,\nev,\neva,\hex,\uex,\bgses} \\
\multicolumn{15}{c}{}   &                                                                                               \tiny{\sses,\bhes,\hes,\nes,\ben,\den,\hno }\\ 
\multicolumn{15}{c}{}   &                                                                                               \tiny{\wnt,\lnf,\enf,\rnv} \\
16164$-$0746 &             & $16$&$19$&$12$ & $-07$&$54$&$03$ & $<$25 &         & 0.021  &  10&2  & s &   25.4 & \tiny{\nen} \\
\end{tabular}  
\end{minipage}
\end{table*}

\begin{table*}
 \centering
 \begin{minipage}{170mm}
 \contcaption{}
 \begin{tabular}{ll r@{:}c@{:}l r@{:}c@{:}l rrlr@{.}llrl}

 \multicolumn{1}{c}{\em IRAS} & & \multicolumn{6}{c}{\bf Position (J2000)} & 
 \multicolumn{1}{c}{\bf OH} & \multicolumn{1}{c}{\bf \h2o} & 
 \multicolumn{1}{c}{} & \multicolumn{2}{c}{\em IRAS} & 
 \multicolumn{1}{c}{\bf Sel.$^a$} & \multicolumn{1}{c}{\bf RMS$^b$} & \\

\multicolumn{1}{c}{\bf Name}  & \multicolumn{1}{l}{\bf Alias} & 
 \multicolumn{3}{c}{\bf RA} & \multicolumn{3}{c}{\bf Dec} & 
 \multicolumn{2}{c}{\bf Peak Flux} & \multicolumn{1}{c}{\bf z} & 
 \multicolumn{2}{c}{\bf 60$\mu$m} & \multicolumn{1}{c}{\bf Crit.} & 
 \multicolumn{1}{c}{\bf Noise} & {\bf References$^{c}$} \\

& & \multicolumn{3}{c}{} & \multicolumn{3}{c}{} & 
 \multicolumn{2}{c}{\bf (mJy)} & \multicolumn{1}{c}{} & 
 \multicolumn{2}{c}{\bf Jy} & & \multicolumn{1}{c}{\bf (mJy)} & \\
\\
16330$-$6820 & ESO 69-6    & $16$&$38$&$13$ & $-68$&$26$&$43$ & $<$35 &         & 0.0456 &   7&2  & z &  121.4 & \tiny{\ssnt} \\
16399$-$0937 &             & $16$&$42$&$40$ & $-09$&$43$&$14$ &    25 &         & 0.0267 &   8&5  & h &   16.8 & \tiny{\ssex,\hno} \\
16504$+$0229 & NGC~6240    & $16$&$52$&$59$ & $+02$&$23$&$59$ & $<$20 &   $<$90 & 0.0245 &  23&5  & sz&   13.4 & \tiny{\sses,\nes,\nen,\bnx} \\
17208$-$0014 &             & $17$&$23$&$21$ & $-00$&$17$&$00$ &   125 &         & 0.0428 &  34&1  &hlsz&  10.6 & \tiny{\bgev,\sses,\hes,\ben,\nen,\men,\bhen} \\
\multicolumn{15}{c}{}   &                                                                                      \tiny{\hno,\wnt,\rnv} \\
17260$-$7622 &             & $17$&$33$&$14$ & $-76$&$24$&$48$ & $<$20 &         & 0.0181 &   4&2  & o &   18.1 & \tiny{\ssnt} \\
17422$-$6437 & IC 4662     & $17$&$47$&$06$ & $-64$&$38$&$25$ & $<$12 &         & 0.0011 &   8&3  & o &   22.1 & \tiny{\ssnt} \\
18093$-$5744 & ESO 140-10  & $18$&$13$&$40$ & $-57$&$43$&$38$ & $<$25 &         & 0.017  &  15&2  & s &   20.9 & \tiny{\ssnt} \\
18293$-$3413 &             & $18$&$32$&$40$ & $-34$&$11$&$26$ & $<$18 &         & 0.018  &  35&3  & sz&   41.2 & \tiny{\nen,\ssnt} \\
18325$-$5926 & Fair 49     & $18$&$36$&$59$ & $-59$&$24$&$09$ & $<$32 &  $<$575 & 0.0192 &   3&2  & o &   20.7 & \tiny{\nen\bnx} \\
18401$-$6225 & ESO 140-43  & $18$&$44$&$47$ & $-62$&$21$&$57$ & $<$35 &  $<$290 & 0.0136 &   2&0  & o &   18.1 & \tiny{\nen,\bnx} \\
18421$-$5049 & ESO 230-10  & $18$&$46$&$02$ & $-50$&$46$&$30$ & $<$20 &         & 0.0177 &   5&1  & o &   15.9 & \tiny{\ssnt} \\
18508$-$7815 & QSO1850-782 & $18$&$58$&$33$ & $-78$&$11$&$49$ &       &         & 0.1618 &   1&1  & o &   11.3 & \\
19115$-$2124 & ESO 593-8   & $19$&$14$&$32$ & $-21$&$19$&$04$ & $<$20 &         & 0.0495 &   6&2  & z &   17.6 & \tiny{\ssnt} \\
19254$-$7245 & Superantennae & $19$&$31$&$21$ & $-72$&$39$&$22$ & $<$40 &       & 0.0615 &   5&2  & z &    3.6 & \tiny{\ssnt} \\
19297$-$0406 &             & $19$&$32$&$21$ & $-04$&$00$&$06$ &       &         & 0.0856 &   7&2  & lz&   14.1 & \\
19393$-$5846 & NGC~6810    & $19$&$43$&$34$ & $-58$&$39$&$21$ & $<$25 &         & 0.006  &  18&1  & s &   20.2 & \tiny{\nen} \\
20087$-$0308 &             & $20$&$11$&$23$ & $-02$&$59$&$54$ &       &         & 0.1033 &   4&6  & lz&   15.6 & \\
20100$-$4156 &             & $20$&$13$&$30$ & $-41$&$47$&$35$ &   200 & $<$3500 & 0.129  &   5&2  &hlz&    4.8 & \tiny{\ssen,\hno,\knl,\ssnt,\rnv,\bnx,\knx} \\
20414$-$1651 &             & $20$&$44$&$17$ & $-16$&$40$&$14$ &       &         & 0.0871 &   4&7  & lz&   11.5 & \\
20491$+$1846 & Z 448-16    & $20$&$51$&$26$ & $+18$&$58$&$08$ & y$^e$ &         & 0.0283 &   2&8  & h &    9.4 & \tiny{\bgen} \\
20550$+$1656 & II Zw 96    & $20$&$57$&$24$ & $+17$&$07$&$40$ &    40 &         & 0.0362 &  13&1  &hsz&    9.2 & \tiny{\bex,\sses,\ben,\hno,\wnt} \\
20551$-$4250 & ESO 286-19  & $20$&$58$&$27$ & $-42$&$39$&$06$ & $<$30 &         & 0.0428 &  12&7  & sz&   11.6 & \tiny{\nen,\ssnt} \\
21130$-$4446 &             & $21$&$34$&$15$ & $-44$&$32$&$43$ & $<$20 &         & 0.0925 &   3&2  & l &   12.2 & \tiny{\ssnt} \\
21219$-$1757 & QSO2121-179 & $21$&$24$&$42$ & $-17$&$44$&$46$ &       &         & 0.113  &   1&1  & o &    8.7 & \\
21330$-$3846 & ESO 343- 13 & $21$&$36$&$11$ & $-38$&$32$&$37$ & $<$35 &         & 0.0191 &   6&8  & o &   12.5 & \tiny{\ssnt} \\
21453$-$3511 & NGC~7130    & $21$&$48$&$19$ & $-34$&$57$&$09$ & $<$20 &  $<$450 & 0.0161 &  16&7  & s &   24.7 & \tiny{\nen,\ssnt,\bnx} \\
22287$-$1917 & ESO 602-25  & $22$&$31$&$25$ & $-19$&$01$&$60$ & $<$25 &         & 0.025  &   5&8  & o &   15.7 & \tiny{\bnt} \\
22467$-$4906 & ESO 239-2   & $22$&$49$&$40$ & $-48$&$50$&$59$ & $<$20 &         & 0.0423 &   6&6  & o &   15.4 & \tiny{\ssnt} \\
22491$-$1808 &             & $22$&$51$&$49$ & $-17$&$52$&$24$ &    15 &         & 0.0777 &   5&5  &hz &   10.5 & \tiny{\bnt,\hno,\knl} \\
23007$+$0836 & NGC~7469    & $23$&$03$&$16$ & $+08$&$52$&$26$ &  $<$8 &   $<$60 & 0.0162 &  27&0  & sz&   15.6 & \tiny{\sses,\nen,\nnv,\bnx} \\
23128$-$5919 & ESO 148-2   & $23$&$15$&$47$ & $-59$&$03$&$17$ & $<$22 &         & 0.0447 &  11&1  & sz&   17.8 & \tiny{\nen,\ssnt} \\
23156$-$4238 & NGC~7582    & $23$&$18$&$23$ & $-42$&$22$&$11$ & $<$32 &  $<$265 & 0.0053 &  48&0  & s &   35.4 & \tiny{\nen,\ssnt,\bnx} \\
23230$-$6926 &             & $23$&$26$&$04$ & $-69$&$10$&$16$ & $<$20 &         & 0.1062 &   3&7  & lz&   13.5 & \tiny{\ssnt} \\
\end{tabular} 

\medskip

NOTES:

(a): The selection parameters satisfied by this source (see Section
2.2). s=S60$>$10; z=S60$\times$z$>$0.3; l=logLIR$>$12; h=known OH
megamaser; w=known \h2o\/ megamaser; o=other {\em IRAS} galaxy.

(b): The quoted rms is after boxcar smoothing of five channels. The data were
also inspected visually with no smoothing, Hanning smoothing and box-car 
smoothing of five and ten channels.

(c): 
\wsf=\cite{wg74};
\wsv=\cite{wg75};
\lss=\cite{ld77};
\dsn=\cite{dl79};
\bet=\cite{bw82};
\get=\cite{gw82};
\bef=\cite{bh84};
\nef=\cite{n84};
\mef=\cite{mg84};
\cef=\cite{ch84};
\nev=\cite{nb85};
\neva=\cite{n85};
\tev=\cite{t85};
\bgev=\cite{bg85};
\hex=\cite{hw86};
\uex=\cite{uc86};
\nex=\cite{nw86};
\cex=\cite{cl86};
\wex=\cite{wg86};
\ccex=\cite{cc86};
\bex=\cite{bg86};
\ssex=\cite{ss86};
\bgses=\cite{bgs87};
\sses=\cite{ss87};
\bhes=\cite{bh87};
\bhhes=\cite{bhh87};
\hmes=\cite{hm87};
\hes=\cite{hg87};
\nes=\cite{nc87};
\mes=\cite{ms87};
\bges=\cite{bg87};
\mee=\cite{mk88};
\nee=\cite{nk88};
\kee=\cite{km88};
\meea=\cite{mb88};
\meeb=\cite{ml88};
\ssen=\cite{ss89};
\ben=\cite{b89};
\den=\cite{dn89};
\nen=\cite{ng89};
\men=\cite{mr89};
\bhen=\cite{bh89};
\bgen=\cite{bg89};
\ken=\cite{km89};
\cno=\cite{cs90};
\hno=\cite{hb90};
\kno=\cite{kp90};
\dno=\cite{dp90};
\knl=\cite{kb91};
\mnt=\cite{mc92};
\bnt=\cite{bh92};
\ssnt=\cite{ss92};
\bnf=\cite{bw94};
\lnf=\cite{ld94};
\enf=\cite{en94};
\vlnv=\cite{vl95};
\rnv=\cite{rf95};
\nnv=\cite{ni95};
\gnx=\cite{gb96};
\bnx=\cite{bw96};
\ggnx=\cite{gg96};
\knx=\cite{ks96};
\gns=\cite{ge97};

(d): These sources were not part of the current survey, but were
observed by Ellingsen et al. \shortcite{en94} and included for
completeness.

(e): These sources have been detected, but only the isotropic luminosity
has been published.
\end{minipage}
\end{table*}

\subsection{Megamaser survey}
All spectra for each source (phase-shifted and smoothed) were
inspected visually for possible peaks in the data. A summary of the
results is shown in Table~\ref{mmmtab}. None of the 87 observed
sources contained detectable 6.7-GHz emission. Assuming a detection
threshold of 5 $\sigma$ this gives a flux-density limit of 15 to
25~mJy for most sources, which is sufficient to detect almost all
known OH and \h2o\ megamasers. The peak flux density of known OH and
\h2o\ megamasers associated with these sources (as well as peak flux
density limits) is also given in Table~\ref{mmmtab}.

\section{Discussion} 
The 6.7-GHz $5_1-6_0~\mbox{A}^+$ transition of methanol is the second
strongest masing transition observed in Galactic sources and is
extremely common, which implies that it is relatively easy to
produce the conditions needed for the molecule to produce maser
emission.  These masers are believed to exist in regions of star
formation only and are closely related to OH masers. Most OH masers sources
also show methanol emission, and vice-versa. Caswell et al.
\shortcite{cv95} have shown that the OH and methanol emission is
generally coincident to within 1 arcsec. This is interpreted as
meaning that the OH and methanol masers require similar physical
conditions.  23 known OH megamaser sources were included in
the sample (see Table~\ref{mmmtab}). Given the close association of
Galactic OH and methanol masers it is surprising that methanol
emission is not seen in any of these sources.

\subsection{NGC 253}
In NGC253, the OH maser emission has a flux density of 0.1~Jy. If the
Galactic methanol/OH maser intensity ratio ($\sim$10) were to apply in
NGC253, then we would expect to observe a 1-Jy methanol maser, which
is ruled out by our observations.

Even if the methanol maser emission does not scale with the OH
emission, the high star-formation rate 
in NGC253 might lead us to expect NGC253 to contain stronger
methanol masers associated with star-formation than those in our own
Galaxy. Our limit of 107~mJy rules out Galactic-type masers 35 times
stronger than those in our own Galaxy.

\subsection{Megamaser survey}
Our sample of megamaser candidates encompasses most of the types of
galaxies in which methanol megamasers might reasonably be expected to
be found. Thus our failure to detect any suggests that 6.7-GHz
methanol megamasers do not exist. This result has immediate
implications for models of OH megamasers.

One such model suggests that OH megamasers represent a large
collection of Galactic-type masers \cite{bw82}.  As methanol is
typically ten times stronger than OH in Galactic sources \cite{cv95}
methanol should be observable with our sensitivity in some, if not
all, OH megamaser galaxies.  Thus these observations rule out this
model of OH megamasers.

We also note that our result rules out another (unpublished) model for
megamasers in which a massive black hole at the galactic nucleus
gravitationally lenses normal masers in star-formation regions on the
far side of the galaxy. Since Galactic OH masers are usually
accompanied by much stronger methanol masers, this model predicts
that we should detect methanol megamasers in a substantial fraction of
the OH megamaser galaxies.

This result has implications for the ``standard'' model of megamaser
emission \cite{n84}, in which low-gain maser amplification of a
background source occurs in molecular clouds within the galactic disc.
Within a megamaser galaxy, the maser path need not be confined to any
one molecular cloud, but instead may encompass many different regions
with widely varying conditions. Only those regions which have the
right abundance, pump conditions, and velocity, will contribute to the
intensity of the maser. However, this process is not critically
dependent on any one condition (e.g. the precise value of optical
depth to the pump), because the maser path will generally sample a
wide range of conditions.

An OH megamaser galaxy has, by virtue of its OH emission, already
demonstrated that it has a high column density of molecular material,
aligned at a suitable velocity against a suitable input source to the
maser amplifier, so that all the conditions for methanol megamaser
emission are already satisfied provided that (a) the methanol
abundance is sufficiently high, and (b) that somewhere along the line
of sight, there exist suitable pumping conditions to invert the
methanol population.

We now consider possible reasons for the absence of methanol
megamasers.

$\bullet$ The methanol abundance may be too low in these galaxies to
support masing on the scale needed for megamaser emission. However,
Henkel et~al. \shortcite{hj87} detected methanol at millimetre
wavelengths towards NGC~253 and IC~342, with abundances similar to
those in our own Galaxy.  This rules out a low methanol abundance as
the reason for the lack of methanol megamasers.

$\bullet$ Methanol may not form on large scales. If the methanol
abundance was not sufficiently high in the molecular clouds which
constitute the maser amplifiers, we would not expect to see megamaser
emission. Galactic methanol masers are seen only in star-formation
regions and the molecules are generally thought to form on the surface
of dust grains, and are released upon the destruction of the grain
\cite{h91}. If there were processes operating within molecular clouds
which depleted the abundance of methanol, then methanol abundances
sufficient for maser action might be confined to star-formation
regions and not exist on the large scales needed to produce
megamasers.

$\bullet$ Appropriate pumping conditions may not exist for methanol
masers to operate on a sufficiently large scale. We consider this less
likely because a) both OH megamasers and (Galactic) methanol masers
are believed to be pumped by IR radiation, b) the widespread nature of
methanol masers in star-formation regions suggests that it is
relatively easy to pump methanol, c) as noted above, it is unlikely
that the appropriate conditions do not occur anywhere along the masing
path. However, we note that the methanol pumping mechanism is not yet
fully understood, and it is possible that it requires special
conditions, such as very high densities.

In this section, we have not considered implications of this result
for \h2o\ megamasers, as the pump mechanisms for OH and methanol
(probably IR) is significantly different from that of \h2o\ megamasers
(collisionally pumped by X-ray-heated gas).

\section{Conclusions}

A survey of 87 {\em IRAS} galaxies for 6.7-GHz methanol masers
detected no emission towards any of the galaxies. Our detection limits
are sufficiently low to detect most known OH and \h2o\ megamasers.
Our sample encompasses most of the types of galaxies in which we might
expect to find methanol megamasers, and the sample size is large
enough that we can reasonably conclude that methanol megamasers do not
exist. This result has two significant implications:

$\bullet$ the absence of methanol megamasers rules out models for OH
megamasers involving large numbers of HII regions similar to those in
our galaxy.

$\bullet$ it suggests that OH megamasers are located in large
molecular clouds in which the abundance of methanol is much lower than
that of OH.

\section{Acknowledgments}
The Australia Telescope is funded by the Commonwealth Government for
operation as a National Facility managed by the CSIRO.  This research
used the SIMBAD data base, operated at CDS, Strasbourg, France and the
NASA/IPAC Extragalactic Database (NED) which is operated by the Jet
Propulsion Laboratory, California Institute of Technology.

\label{lastpage}

\end{document}